\documentclass[11pt,a4paper]{article}

\usepackage{jheppub}
\usepackage{amssymb,enumerate}
\usepackage{amsmath}
\usepackage{axodraw}
\usepackage{bbm}
\usepackage{url}
\usepackage{graphics}
\usepackage{xspace}
\usepackage{epsfig}
\usepackage{subfigure}
\usepackage{booktabs}

\graphicspath{ {figures/} }

\allowdisplaybreaks[2]

\newcommand{\POWHEG}{{\tt POWHEG}}
\newcommand{\POWHEGBOX}{{\tt POWHEG\;BOX}}
\newcommand{\PYTHIA}{{\tt PYTHIA}}
\newcommand{\POWHEGpPYTHIA}{{\tt POWHEG+PYTHIA}}
\newcommand{\muf}{\mu_\mathrm{F}}
\newcommand{\mur}{\mu_\mathrm{R}}
\newcommand{\mr}{\mathrm}
\newcommand{\slsl}{\tilde{\ell}^+\tilde{\ell}^-}
\newcommand{\slslj}{\tilde{\ell}^+\tilde{\ell}^-+\mr{jet}}

\makeatletter
\renewcommand\@fpheader{\hfill \parbox{2.4cm}{MITP/14-075 \\ DESY 14-184}}
\renewcommand\@journal{}
\makeatother

\title{Slepton pair production in association with a jet: NLO-QCD corrections and parton-shower effects}

\author[a]{Barbara~J\"ager,}
\author[b]{Andreas~von~Manteuffel,}
\author[b,c]{Stephan~Thier}  

\affiliation[a]{
 Institute for Theoretical Physics,
 University of T\"ubingen,
 Auf der Morgenstelle 14,
 72076 T\"ubingen, Germany
}
\affiliation[b]{
  PRISMA Cluster of Excellence,
  Institute of Physics,
  Johannes Gutenberg University, 
  55099 Mainz, Germany
}
\affiliation[c]{
  II. Institute for Theoretical Physics,
  Hamburg University,
  22761 Hamburg, Germany
}

\emailAdd{jaeger@itp.uni-tuebingen.de}
\emailAdd{manteuffel@uni-mainz.de}
\emailAdd{stephan.christoph.thier@desy.de}

\abstract{
We present a calculation of the next-to-leading order QCD corrections to slepton pair production in association with a jet at the LHC together with their implementation in the \POWHEGBOX{}. For the simulation of parton-shower effects and the decays of the sleptons we employ the multi-purpose Monte-Carlo program {\tt PYTHIA}. 
We discuss the impact of next-to-leading order QCD corrections on experimentally accessible distributions and illustrate how the parton shower can modify observables that are sensitive to QCD radiation effects.
Having full control on the hard jet in the process, we provide precise predictions
also for monojet analyses.
}

\begin{document}

\maketitle

%
\section{Introduction}
With the start-up of the CERN Large Hadron Collider (LHC) particle physics has entered a new era, culminating in the discovery of the Higgs boson~\cite{Aad:2012tfa,Chatrchyan:2012ufa}. Nonetheless, we are left with a plethora of open questions pointing to the necessity of extending the Standard Model (SM) of elementary particles. A theoretically particularly appealing approach is the construction of supersymmetric (SUSY) theories that predict the existence of new particles which, before SUSY breaking, merely differ in their spin quantum numbers from the SM partners they are associated with.

As of yet, the ATLAS and CMS collaborations are providing severe exclusion limits on strongly interacting supersymmetric particles (see, e.g., Refs.~\cite{ATLAS-susy,CMS-susy}). Because of smaller production cross sections, weakly interacting supersymmetric particles are more difficult to access, but have recently gained increasing attention by both experimental collaborations \cite{Aad:2012pxa,Aad:2012hba,Chatrchyan:2012pka}. 
In the context of the minimal supersymmetric extension of the Standard Model (MSSM) much effort has also been made by theorists to provide precise predictions for the production of these color neutral particles.
Perturbative calculations for the pair production of the scalar partners of the leptons, the sleptons, for instance, have been ever refined during the last decades. 
Next-to-leading order (NLO) QCD and SUSY-QCD (SQCD) corrections to slepton pair production processes at hadron colliders\footnote{
Precise determinations of the slepton properties could be performed at a future
$e^+e^-$ linear collider~\cite{Martyn:1999tc,Freitas:2003yp,Freitas:2004kh}, and, with
restrictions, at a muon collider~\cite{Freitas:2011ti}.}
have first been computed in \cite{Baer:1997nh} and \cite{Beenakker:1999xh}, respectively. The latter calculation is publicly available in the format of the computer package {\tt PROSPINO}~\cite{Beenakker:1996ed}. Resummation effects have been considered in Refs.~\cite{Bozzi:2006fw,Bozzi:2007qr,Bozzi:2007tea,Broggio:2011bd}.
In Refs.~\cite{FridmanRojas:2012yh} and \cite{Jager:2012hd}, NLO-(S)QCD corrections to slepton pair production processes have been matched with {\tt HERWIG++}~\cite{Marchesini:1991ch,Corcella:2000bw} and {\tt PYTHIA}~\cite{Sjostrand:2006za}, respectively, making use of the {\tt POWHEG} method~\cite{Nason:2004rx,Frixione:2007vw}, an approach that allows to combine fixed-order perturbative calculations with parton-shower programs in a well-defined manner. 

While parton-shower programs are capable of simulating the emission of soft and/or collinear partons in a hard-scattering event, they are not designed to account for extra hard emissions. If processes with extra hard jets in the final state, as observed frequently at the LHC, are to be described realistically, the hard-scattering amplitudes themselves have to account for these jets.
This is particularly important for the study of monojet signatures in scenarios where the decay products of the sleptons are difficult to detect.
In this work we therefore provide an explicit NLO-SQCD calculation for slepton pair production in association with a hard jet. Moreover, we describe the implementation of our calculation in the \POWHEGBOX{}~\cite{Alioli:2010xd}, a repository that contains all process-independent ingredients of the \POWHEG{} method. We discuss the implications of the NLO corrections as well as of the parton shower on experimentally accessible distributions. Comparison to previous work performed in the same framework allows us to demonstrate how the description of the hard jet gains from explicit matrix elements for the full hard-scattering process in 
the relevant kinematic domains.


\section{Technical details of the calculation}
\label{sec:calc}
Our calculation of the NLO-(S)QCD corrections to slepton pair production in association with a jet proceeds along similar lines as our previous calculation \cite{Jager:2012hd} for slepton pair production at the LHC in the framework of the MSSM. 

At leading order, we encounter the annihilation of a massless quark-antiquark pair into an intermediate $Z$ boson or photon that in turn decays into a slepton pair, accompanied by a gluon emitted from either of the incoming partons, as shown in Fig.~\ref{fig:feynman_QCD} (a). In addition to the process $q \bar q \to \slsl g$, crossing-related processes with a quark or antiquark and a gluon in the initial state, such as $q g \to \slsl q$ or $\bar q g \to \slsl \bar q$, occur. 
\begin{figure}
\begin{center}
\begin{picture}(400,175)(0,0)
\SetOffset(20,100)
\SetScale{0.35}
\Vertex(100,60){2}
\Vertex(100,140){2}
\Vertex(210,140){2}
\ArrowLine(40,140)(100,140)
\ArrowLine(100,140)(100,60)
\ArrowLine(100,60)(40,60)
\Photon(100,140)(210,140){4}{8.5}
\DashArrowLine(210,140)(260,180){4}
\DashArrowLine(260,100)(210,140){4}
\Gluon(100,60)(260,60){5}{11}
\SetScale{1}
\SetOffset(0,0)
\SetOffset(160,100)
\SetScale{0.35}
\Vertex(110,30){2}
\Vertex(110,60){2}
\Vertex(110,140){2}
\Vertex(220,140){2}
\ArrowLine(30,140)(110,140)
\ArrowLine(110,140)(110,60)
\ArrowLine(110,60)(110,30)
\ArrowLine(110,30)(30,30)
\Photon(110,140)(220,140){4}{9.5}
\DashArrowLine(220,140)(270,180){4}
\DashArrowLine(270,100)(220,140){4}
\Gluon(110,60)(270,60){5}{12}
\Gluon(110,30)(270,30){5}{12}
\SetScale{1}
\SetOffset(0,0)
\SetOffset(300,100)
\SetScale{0.35}
\Vertex(110,30){2}
\Vertex(110,60){2}
\Vertex(110,140){2}
\Vertex(220,140){2}
\ArrowLine(30,140)(110,140)
\ArrowLine(110,140)(110,60)
\ArrowLine(110,60)(270,60)
\ArrowLine(30,30)(110,30)
\ArrowLine(110,30)(270,30)
\Photon(110,140)(220,140){4}{9.5}
\DashArrowLine(220,140)(270,180){4}
\DashArrowLine(270,100)(220,140){4}
\Gluon(110,60)(110,30){5}{2}
\SetScale{1}
\SetOffset(0,0)
\SetOffset(20,0)
\SetScale{0.35}
\Vertex(100,20){2}
\Vertex(100,50){2}
\Vertex(100,110){2}
\Vertex(100,140){2}
\Vertex(210,140){2}
\ArrowLine(40,140)(100,140)
\ArrowLine(100,140)(100,110)
\ArrowLine(100,110)(100,50)
\GlueArc(100,80)(30,-90,90){5}{5}
\ArrowLine(100,50)(100,20)
\ArrowLine(100,20)(40,20)
\Photon(100,140)(210,140){4}{8.5}
\DashArrowLine(210,140)(260,180){4}
\DashArrowLine(260,100)(210,140){4}
\Gluon(100,20)(260,20){5}{11}
\SetScale{1}
\SetOffset(0,0)
\SetOffset(160,0)
\SetScale{0.35}
\Vertex(60,90){2}
\Vertex(110,90){2}
\Vertex(180,130){2}
\Vertex(230,130){2}
\Vertex(180,50){2}
\ArrowLine(20,130)(60,90)
\ArrowLine(60,90)(20,50)
\Gluon(60,90)(110,90){5}{3}
\ArrowLine(110,90)(180,130)
\ArrowLine(180,130)(180,50)
\ArrowLine(180,50)(110,90)
\Photon(180,130)(230,130){4}{3.5}
\DashArrowLine(230,130)(280,170){4}
\DashArrowLine(280,90)(230,130){4}
\Gluon(180,50)(280,50){5}{6}
\SetScale{1}
\SetOffset(0,0)
\SetOffset(300,0)
\SetScale{0.35}
\Vertex(80,60){2}
\Vertex(160,60){2}
\Vertex(160,140){2}
\Vertex(80,140){2}
\Vertex(230,140){2}
\ArrowLine(20,140)(80,140)
\ArrowLine(80,60)(20,60)
\ArrowLine(160,60)(80,60)
\ArrowLine(160,140)(160,60)
\ArrowLine(80,140)(160,140)
\Gluon(80,60)(80,140){5}{4}
\Photon(160,140)(230,140){4}{5.5}
\DashArrowLine(230,140)(280,180){4}
\DashArrowLine(280,100)(230,140){4}
\Gluon(160,60)(280,60){5}{8}
\SetScale{1}
\SetOffset(0,0)
\Text(10,150)[]{(a1)}
\Text(30,150)[]{\small{$q$}}
\Text(30,122)[]{\small{$\bar{q}$}}
\Text(75,158)[]{\small{$\gamma/Z$}}
\Text(115,165)[l]{\small{$\tilde{\ell}^-$}}
\Text(115,137)[l]{\small{$\tilde{\ell}^+$}}
\Text(115,120)[l]{\small{$g$}}
\Text(150,150)[]{(b1)}
\Text(290,150)[]{(b2)}
\Text(10,50)[]{(c1)}
\Text(150,50)[]{(c2)}
\Text(290,50)[]{(c3)}
\end{picture}
\caption{\label{fig:feynman_QCD}
Representative Feynman diagrams for slepton pair production in association with a jet at leading order (a1), with an additional real parton (b1)-(b2), and with virtual SM QCD corrections (c1)-(c3).
}
\end{center}
\end{figure}
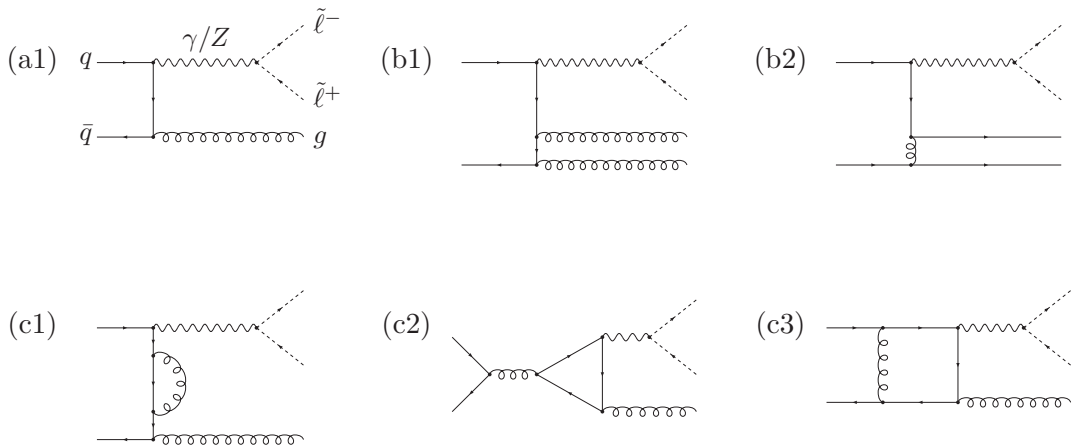
\begin{figure}
\begin{center}
\begin{picture}(400,175)(0,0)
\SetOffset(20,100)
\SetScale{0.35}
\Vertex(100,20){2}
\Vertex(100,50){2}
\Vertex(100,110){2}
\Vertex(100,140){2}
\Vertex(210,140){2}
\ArrowLine(40,140)(100,140)
\ArrowLine(100,140)(100,110)
\DashArrowLine(100,110)(100,50){4}
\GlueArc(100,80)(30,-90,90){5}{5}
\CArc(100,80)(30,-90,90)
\ArrowLine(100,50)(100,20)
\ArrowLine(100,20)(40,20)
\Photon(100,140)(210,140){4}{8.5}
\DashArrowLine(210,140)(260,180){4}
\DashArrowLine(260,100)(210,140){4}
\Gluon(100,20)(260,20){5}{11}
\SetScale{1}
\SetOffset(0,0)
\SetOffset(160,100)
\SetScale{0.35}
\Vertex(100,60){2}
\Vertex(100,140){2}
\Vertex(170,100){2}
\Vertex(230,140){2}
\ArrowLine(40,140)(100,140)
\ArrowLine(100,60)(40,60)
\DashArrowLine(100,140)(170,100){4}
\DashArrowLine(170,100)(100,60){4}
\Gluon(100,140)(100,60){-5}{4}
\Line(100,140)(100,60)
\Photon(170,100)(230,140){4}{5.5}
\DashArrowLine(230,140)(280,180){4}
\DashArrowLine(280,100)(230,140){4}
\Gluon(170,100)(280,20){5}{10}
\SetScale{1}
\SetOffset(0,0)
\SetOffset(300,100)
\SetScale{0.35}
\Vertex(80,60){2}
\Vertex(160,60){2}
\Vertex(160,140){2}
\Vertex(80,140){2}
\Vertex(230,140){2}
\ArrowLine(20,140)(80,140)
\ArrowLine(80,60)(20,60)
\DashArrowLine(160,60)(80,60){4}
\DashArrowLine(160,140)(160,60){4}
\DashArrowLine(80,140)(160,140){4}
\Gluon(80,60)(80,140){5}{4}
\Line(80,60)(80,140)
\Photon(160,140)(230,140){4}{5.5}
\DashArrowLine(230,140)(280,180){4}
\DashArrowLine(280,100)(230,140){4}
\Gluon(160,60)(280,60){5}{8}
\SetScale{1}
\SetOffset(0,0)
\SetOffset(20,0)
\SetScale{0.35}
\Vertex(60,90){2}
\Vertex(110,90){2}
\Vertex(180,130){2}
\Vertex(230,130){2}
\Vertex(180,50){2}
\ArrowLine(20,130)(60,90)
\ArrowLine(60,90)(20,50)
\Gluon(60,90)(110,90){5}{3}
\DashArrowLine(110,90)(180,130){4}
\DashArrowLine(180,130)(180,50){4}
\DashArrowLine(180,50)(110,90){4}
\Photon(180,130)(230,130){4}{3.5}
\DashArrowLine(230,130)(280,170){4}
\DashArrowLine(280,90)(230,130){4}
\Gluon(180,50)(280,50){5}{6}
\SetScale{1}
\SetOffset(0,0)
\SetOffset(160,0)
\SetScale{0.35}
\Vertex(80,60){2}
\Vertex(160,60){2}
\Vertex(160,140){2}
\Vertex(80,140){2}
\Vertex(230,140){2}
\ArrowLine(20,140)(80,140)
\ArrowLine(80,60)(20,60)
\DashArrowLine(160,60)(80,60){4}
\DashArrowLine(160,140)(160,60){4}
\DashArrowLine(80,140)(160,140){4}
\Gluon(80,60)(80,140){5}{4}
\Line(80,60)(80,140)
\DashLine(160,140)(230,140){4}
\DashArrowLine(230,140)(280,180){4}
\DashArrowLine(280,100)(230,140){4}
\Gluon(160,60)(280,60){5}{8}
\SetScale{1}
\SetOffset(0,0)
\SetOffset(300,0)
\SetScale{0.35}
\Vertex(120,60){2}
\Vertex(200,60){2}
\Vertex(200,140){2}
\Vertex(120,140){2}
\Vertex(200,140){2}
\ArrowLine(50,140)(120,140)
\ArrowLine(120,60)(50,60)
\DashArrowLine(200,60)(120,60){4}
\DashArrowLine(200,140)(200,60){4}
\DashArrowLine(120,140)(200,140){4}
\Gluon(120,60)(120,140){5}{4}
\Line(120,60)(120,140)
\DashArrowLine(200,140)(250,180){4}
\DashArrowLine(250,100)(200,140){4}
\Gluon(200,60)(250,60){5}{3}
\SetScale{1}
\SetOffset(0,0)
\Text(10,150)[]{(a1)}
\Text(30,150)[]{\small{$q$}}
\Text(30,108)[]{\small{$\bar{q}$}}
\Text(47,129)[l]{\small{$\tilde{q}$}}
\Text(71,129)[l]{\small{$\tilde{g}$}}
\Text(75,158)[]{\small{$\gamma/Z$}}
\Text(115,165)[l]{\small{$\tilde{\ell}^-$}}
\Text(115,137)[l]{\small{$\tilde{\ell}^+$}}
\Text(115,108)[l]{\small{$g$}}
\Text(150,150)[]{(a2)}
\Text(205,150)[l]{\small{$\tilde{q}$}}
\Text(205,121)[l]{\small{$\tilde{q}$}}
\Text(185,135)[l]{\small{$\tilde{g}$}}
\Text(290,150)[]{(a3)}
\Text(340,158)[l]{\small{$\tilde{q}$}}
\Text(360,135)[l]{\small{$\tilde{q}$}}
\Text(340,113)[l]{\small{$\tilde{q}$}}
\Text(318,135)[l]{\small{$\tilde{g}$}}
\Text(10,50)[]{(a4)}
\Text(68,47)[l]{\small{$\tilde{q}$}}
\Text(87,32)[l]{\small{$\tilde{q}$}}
\Text(68,17)[l]{\small{$\tilde{q}$}}
\Text(150,50)[]{(b1)}
\Text(200,58)[l]{\small{$\tilde{q}$}}
\Text(220,35)[l]{\small{$\tilde{q}$}}
\Text(200,13)[l]{\small{$\tilde{q}$}}
\Text(178,35)[l]{\small{$\tilde{g}$}}
\Text(232,58)[]{\small{$h^0/H^0$}}
\Text(290,50)[]{(c1)}
\Text(354,58)[l]{\small{$\tilde{q}$}}
\Text(374,35)[l]{\small{$\tilde{q}$}}
\Text(354,13)[l]{\small{$\tilde{q}$}}
\Text(332,35)[l]{\small{$\tilde{g}$}}
\end{picture}
\caption{\label{fig:feynman_SQCD}
Representative Feynman diagrams for virtual corrections with an intermediate $Z$ boson or photon and supersymmetric particles in the loop (a1)-(a4), with an intermediate Higgs boson (b1), or with a squark-squark-slepton-slepton vertex (c1).
}
\end{center}
\end{figure}
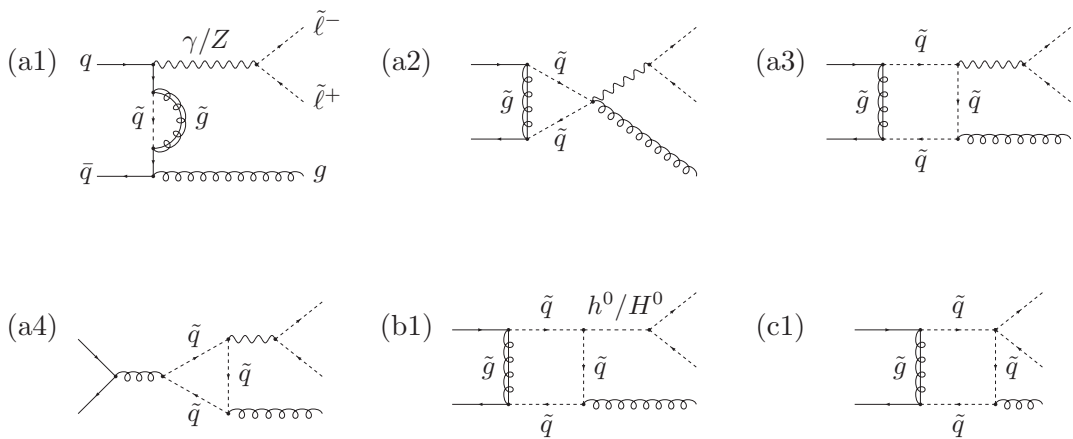

The real-emission contributions comprise scattering processes at order $\alpha_s^2\alpha^2$ with a slepton pair and two partons in the final state, i.~e.\ subprocesses of the type $q\bar q\to\slsl q'\bar q'$, $q\bar q\to\slsl q\bar q$, $q\bar q\to\slsl gg$, and all crossing-related reactions, see Fig.~\ref{fig:feynman_QCD} (b) for two examples.
In order to obtain the respective scattering amplitudes for the Born and the real-emission contributions in a format that can easily be processed by the \POWHEGBOX{}, we made use of the build tool based on {\tt MadGraph~4}~\cite{Murayama:1992gi,Stelzer:1994ta,Alwall:2007st} that was first applied in Ref.~\cite{Campbell:2012am} and is now provided with the public version of the program repository. The default version of the \POWHEGBOX{} is constrained to the Standard Model, but the code can be extended in a straightforward manner to processes involving weakly interacting\footnote{Let us note here that a similar extension to processes involving squarks or gluinos would require special care~\cite{Gavin:2013kga}.}
supersymmetric particles in the context of the MSSM with the help of {\tt SMadGraph}~\cite{Cho:2006sx}.
The suitably adapted build tool then also provides us with the color- and spin-correlated Born amplitudes that are needed for the construction of the counterterms for IR singular configurations in the framework of the Frixione-Kunszt-Signer subtraction formalism~\cite{Frixione:1995ms}. 

The calculation of the virtual (S)QCD corrections is performed with the methods of Ref.~\cite{Jager:2012hd}, with appropriate refinements and extensions to account for the more involved structure of slepton pair production with an extra parton in the final state. 
Throughout, we are using conventional dimensional regularization to handle ultraviolet (UV) and infrared (IR) divergences in a well-defined manner. For the partonic subprocesses under scrutiny, no supersymmetry restoring counterterms are required~\cite{Hollik:2001cz}. 
UV divergences are absorbed by a proper renormalization procedure for quark- and gluon fields
as well as the strong coupling constant~$\alpha_s$ with five active massless flavors.
We work in the on-shell scheme for the renormalization of the quark- and gluon
wavefunctions.
For the renormalization of $\alpha_s$ we employ the $\overline{\mr{MS}}$ scheme,
modified to decouple the top quark\cite{Collins:1978wz,Nason:1989zy} and heavy SUSY
particles~\cite{Berge:2007dz}.
As a consequence, heavy-particle contributions to the counter terms for our
calculation enter only through the quark wave function renormalization.

The virtual corrections can be split into four different groups.
The first two groups resemble the loop corrections to the
Drell-Yan plus jet process with an intermediate $Z$ boson or photon,
either with SM particles,
see Fig.~\ref{fig:feynman_QCD}~(c),
or with supersymmetric particles in the loops,
see Fig.~\ref{fig:feynman_SQCD}~(a), respectively.
In the third group the sleptons originate from one of the MSSM Higgs bosons,
see Fig.~\ref{fig:feynman_SQCD}~(b).
In the fourth group the sleptons originate from a four-particle vertex of two squarks and two sleptons,
see Fig.~\ref{fig:feynman_SQCD}~(c).
Since we employ the Feynman-'t\;Hooft gauge for all gauge fields, the $Z$ boson
exchange diagrams are understood to be supplemented by corresponding Goldstone contributions.
In the simplest case of same-mass sleptons and absence of mixing in the squark sector,
various subsets of these diagrams such as Goldstone contributions, closed squark loop
contributions and contributions with a squark-squark-gluon-$Z$ vertex vanish,
either diagram by diagram or in the sum.

Individual diagrams with a $Z$ boson coupling to a closed fermion loop, see Fig.~\ref{fig:feynman_QCD}~(c2), exhibit the
Adler-Bell-Jackiw axial anomaloy~\cite{Adler:1969gk,Bell:1969ts}.
To calculate its remnant in the full amplitude due to the finite top quark mass
we employ two different schemes for $\gamma_5$ in $D \neq 4$ dimensions,
that were suggested in Refs.~\cite{Kreimer:1989ke,Korner:1991sx} and \cite{Larin:1993tq},
respectively, and find complete agreement.

For diagrams with Drell-Yan plus jet like structure, we apply the decomposition of the process $q + \bar{q} \to \gamma^*/Z^* + g$ into Lorentz structures, which we employed for the real-emission contributions to slepton pair production in Ref.~\cite{Jager:2012hd} already. Projectors constructed from this decomposition are applied to process Feynman diagrams generated with {\tt QGRAF}~\cite{Nogueira:1991ex} using Feynman rules from Ref.~\cite{Rosiek:1989rs} and in-house developed {\tt FORM}~\cite{Vermaseren:2000nd,Kuipers:2012rf} scripts. The loop integrals obtained in this calculation are reduced with {\tt Reduze\;2}~\cite{Studerus:2009ye,vonManteuffel:2012yz,ginac,fermat}, yielding expressions that contain only scalar master integrals which can be evaluated numerically with the {\tt QCDloop} library~\cite{Ellis:2007qk,vanOldenborgh:1990yc}. A second calculation of the Drell-Yan plus jet like virtual corrections based on {\tt FeynArts}~\cite{Hahn:2000kx}, {\tt FormCalc}~\cite{Hahn:1998yk,Hahn:2006zy}, and {\tt LoopTools}~\cite{Hahn:1998yk,vanOldenborgh:1990yc} provides 
an independent check for these contributions.
For our {\tt FormCalc} calculation, we carefully implement alternative routines
for the evaluation of fermion traces according to our treatment of $\gamma_5$.
We find complete agreement between the two calculations.
Virtual corrections featuring Goldstone bosons, Higgs bosons, or squark-squark-slepton-slepton vertices, are implemented in our code based on matrix elements obtained with {\tt FeynArts}, {\tt FormCalc}, and appropriate integral reduction formulae~\cite{Denner:2002ii}.

For the parameterization of the phase space, we adapt the implementation of Ref.~\cite{Alioli:2011as} that was originally developed for the related case of $t\bar t j$ production at the LHC in the  \POWHEGBOX{}. 
The inclusive Born cross section for slepton pair production in association with a jet is singular when the final-state parton becomes soft or collinear to an incoming parton. Once realistic acceptance cuts are imposed on the jet, such contributions are irrelevant for phenomenological applications. However, they spoil the efficiency of the program, if not handled with care. The \POWHEGBOX{} offers two approaches for dealing with singular Born configurations: generation cuts that avoid the population of unwanted regions in phase space from the beginning, and a so-called Born-suppression factor that dampens contributions from singular regions of phase space. In order to ensure our phenomenological results do not depend on the selected procedure, we ran the code with both options. After applying a realistic transverse-momentum cut of $p_\mr{T}^\mr{jet1} >20$~GeV on the hardest jet at analysis level  we found identical results for a setup with a generation cut of 
$p_{\mr{T},i}^\mr{gen} > 10~\mr{GeV} 
$
on the final-state parton $i$ of the underlying Born configuration and a setup with a Born suppression factor of the form
\begin{equation}
\label{eq:bsupp}
F(\Phi_n) = \frac{ p_{\mr{T},i}^2}{p_{\mr{T},i}^2+\Lambda^2}\,,
\end{equation}
with $\Lambda= 10~\mr{GeV} $. For the results presented in Sec.~\ref{sec:pheno}, we employ the Born suppression factor of Eq.~(\ref{eq:bsupp}) with $\Lambda = 10$~GeV together with a mild generation cut of $p_{\mr{T},i}^\mr{gen} > 1~\mr{GeV}$.

While our calculation represents the first complete NLO-SQCD calculation for slepton pair production in association with a jet, the related case of lepton-pair production with an associated jet in NLO QCD has been considered long ago and is available in the \POWHEGBOX{}~\cite{Alioli:2010qp}.
We compare the virtual corrections in this implementation of $pp\to \ell^+\ell^-+\mr{jet}+X$ to the corresponding terms in an adapted version of our code.
Replacing the sleptons of our calculation with leptons, adjusting couplings and input parameters, and selecting appropriate diagrams, we are able to reproduce the results of this code at representative phase-space points.
Since our setup encapsulates the (renormalized) loop contributions in scalar coefficient functions which are independent of the decay of the photon or $Z$ boson,
this comparison provides a very direct check on our calculation of the virtual corrections and their implementation in the \POWHEGBOX{}.

Additional checks are based on the observation that 
the Born amplitudes for $pp\to\slslj+X$ equal the real emission amplitudes of the slepton pair production process $pp\to\slsl+X$. A comparison of our leading-order matrix elements for $pp\to\slslj+X$ with the real-emission amplitudes of \cite{Jager:2012hd}  shows excellent agreement. 
Moreover, for differential distributions of the hardest jet we find full agreement between the two codes ran at leading order and NLO accuracy, respectively, after realistic selection cuts. This provides a powerful test on the phase space integration and the overall normalization of the new code.


\section{Phenomenological results and discussion}
\label{sec:pheno}
Our implementation of slepton pair production in association with a jet in the \POWHEGBOX{} will be made publicly available at {\tt http://powhegbox.mib.infn.it/}. Together with the code we provide a documentation with instructions and  recommended technical parameters for running the program. The interested reader is free to use the default version of the code including routines for a phenomenological analysis, or to adapt input parameters, histograms, and selection cuts to his own needs. In order to demonstrate the capability of the code, here we present 
results for some phenomenologically interesting setups.

We consider proton-proton collisions at the LHC with a center-of-mass energy of $\sqrt{s}=14$~TeV. For the parton-distribution functions of the proton we use the NLO-QCD set of the MSTW2008 parameterization~\cite{Martin:2009iq}, as implemented in the {\tt LHAPDF} library~\cite{Whalley:2005nh}.  Factorization and renormalization scales are set to $\mu_R=\mu_F= \mu_0$ with $\mu_0=2 \,m_{\tilde{\ell}}$, unless explicitly stated otherwise. Statistical uncertainties are negligible for all results presented here.

All required SM and MSSM parameters are provided in a file complying with the SUSY Les Houches Accord ({\tt SLHA})~\cite{Skands:2003cj,Allanach:2008qq}. They are processed by routines based on {\tt MadGraph~4}, which calculate the dependent parameters and all particle couplings.
As electroweak input parameters we are using the mass of the $Z$~boson, $m_Z=91.1876$~GeV, the electromagnetic coupling, $\alpha\left(m_Z\right) = 1/127.944$, and the Fermi constant, $G_\mr{F} = 1.1663787\times 10^{-5}~\mr{GeV}^{-2}$. The top quark mass is set to $m_t=173.07$~GeV, all other quark masses are neglected.

Our {\em default setup} for $pp\to \slslj+X$ features selectrons or smuons with a mass
\begin{equation}
m_{\tilde{\ell}} = 350\text{~GeV}
\end{equation}
that lies above current exclusion limits by ATLAS~\cite{Aad:2014vma} and CMS~\cite{Khachatryan:2014qwa} irrespective of the masses of potential decay products. We do not consider sleptons of the third generation but, for our representative analysis, restrict ourselves to electrically charged left-handed sleptons of one of the first two generations (no sum) without slepton mixing.
The mass of the lightest neutralino $\tilde{\chi}_0^1$ is taken as
\begin{equation}
m_{\tilde{\chi}_1^0}=100\text{~GeV}\,,
\end{equation}
while we assume the other neutralinos and the charginos to be heavier than the sleptons. In this scenario, sleptons exclusively decay into a lepton and the lightest neutralino. We simulate these decays as well as parton-shower effects with the help of {\tt PYTHIA 6.4.25}. Throughout, we switch off QED radiation, underlying event and hadronization effects.

Partons in the final state are recombined into jets with the help of the anti-$k_T$ algorithm of Ref.~\cite{Cacciari:2008gp}, as implemented in the {\tt FASTJET} package~\cite{Cacciari:2005hq,Cacciari:2011ma}, with a resolution parameter of $R=0.4$. 
For our numerical studies we require the presence of at least one jet with
\begin{equation}
\label{cuts:basic}
p_\mr{T}^\mr{jet1}>20~\mr{GeV}\,,\quad
|y^\mr{jet1}|<4.5\,.
\end{equation}

In the setting described above we compute the LO cross section for $pp\to \slslj+X$ and find
\begin{equation}
\sigma_{\text{LO}}\big(\tilde{\ell}\tilde{\ell}j\big) = 1.624~\text{fb}\,.
\end{equation}
Based on the experience from slepton pair production~\cite{Jager:2012hd} and Drell-Yan plus jet production~\cite{Gavin:2011wn}, where virtual corrections with supersymmetric particles in the loops were found to be very small compared to the other contributions, we test the impact of different groups of virtual corrections separately.
Considering just Drell-Yan plus jet like SM contributions we find for the NLO cross section
\begin{equation}
\sigma_{\text{NLO,DY(SM)}}\big(\tilde{\ell}\tilde{\ell}j\big)= 1.826~\text{fb}\,.
\end{equation}
Including also supersymmetric Drell-Yan plus jet like corrections, we find that
even for a light common squark mass of $m_{\tilde{q}} =500$~GeV and a light gluino mass of $m_{\tilde{g}} =700$~GeV
the cross section changes only slightly,
$\sigma_{\text{NLO,DY(SM+SUSY)}}\big(\tilde{\ell}\tilde{\ell}j\big) = 1.847~\text{fb}$.
The remaining corrections with either a Higgs boson of mass $m_{h^0} = 126$~GeV or $m_{H^0} = 400$~GeV
or with a squark-squark-slepton-slepton vertex are found to be completely negligible
at this level of accuracy.
Consequently, we neglect such contributions in our phenomenological study, only the virtual corrections generated by diagrams with QCD loops, see Fig.~\ref{fig:feynman_QCD} (c), are included. In this way, we 
reduce both the number of free parameters in our setup and the execution time of our code.

\begin{figure}[t]
\centerline{
\includegraphics[trim=17mm 0mm 0mm 0mm,height=0.35\textheight,clip]{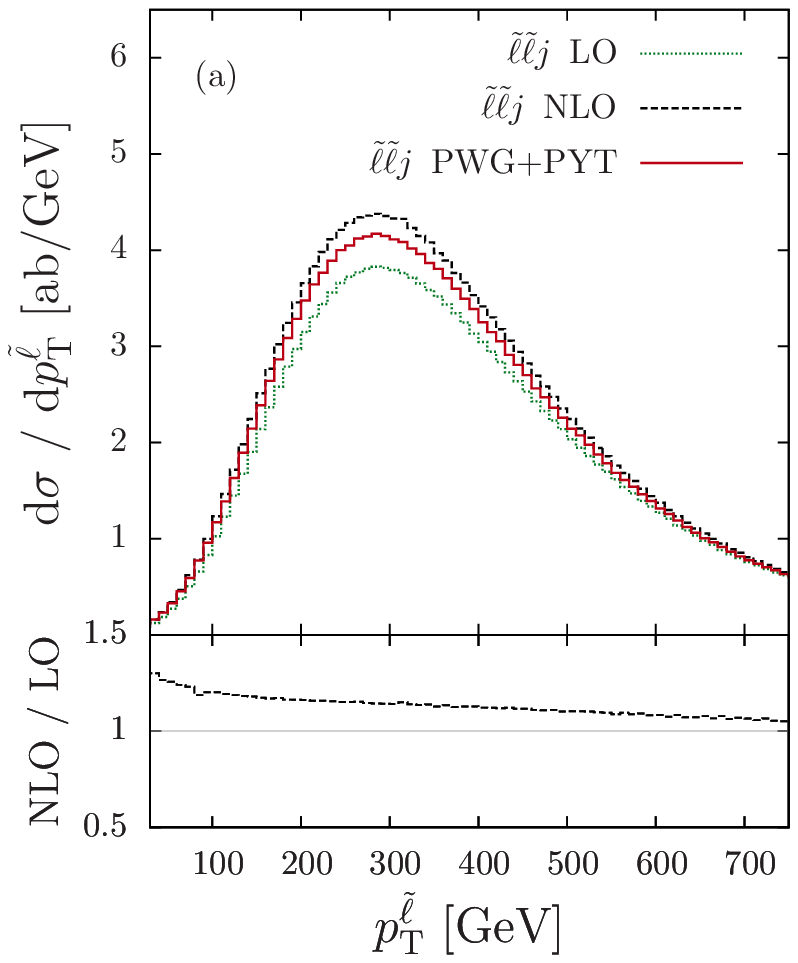}
\includegraphics[trim=12mm 0mm 0mm 0mm,height=0.35\textheight,clip]{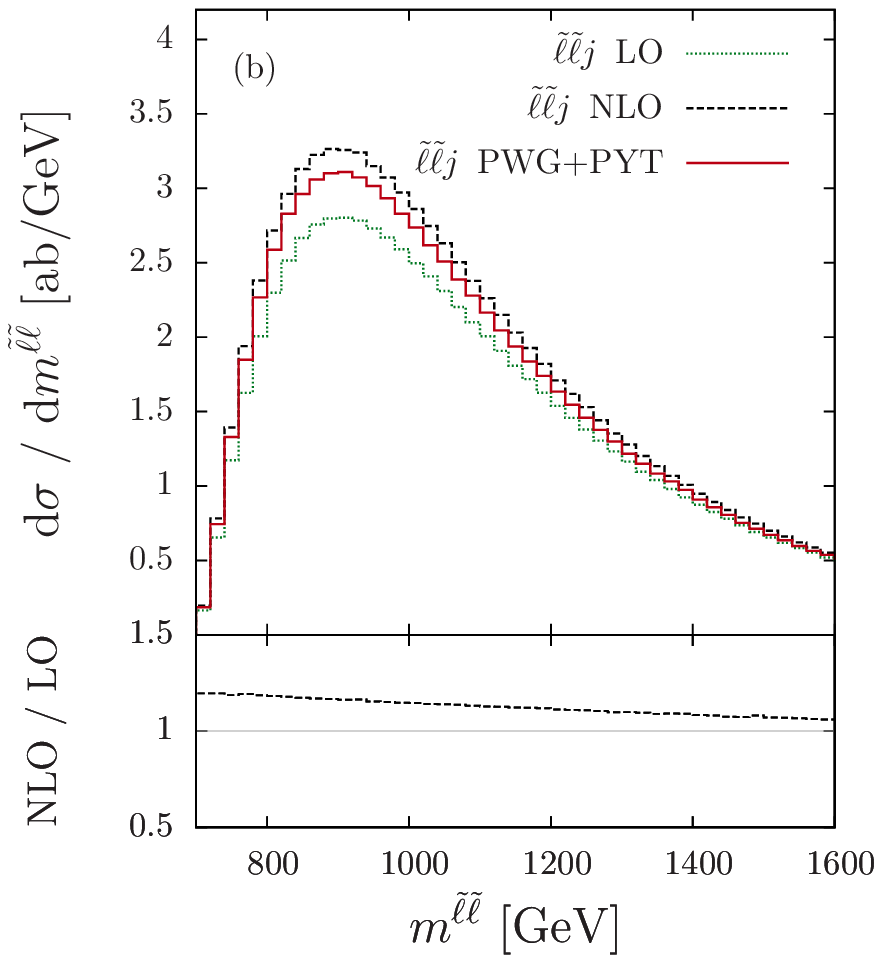}
}
\caption{\label{fig:slsl}
Transverse momentum of the negatively charged slepton~(a) and invariant-mass distribution of the slepton pair~(b) at LO (green dotted), NLO-QCD (black dashed), and with \POWHEGpPYTHIA{} (red solid) in our default setup for $pp\to \slslj+X$.}
\end{figure}

Distributions related to the sleptons receive sizable corrections compared
to LO  results when NLO contributions are included, and the parton shower gives rise
to further noticable effects.
In Fig.~\ref{fig:slsl}
this is illustrated for the transverse momentum of the negatively charged slepton and the invariant mass distribution of the slepton pair in $pp\to \slslj+X$. 

\begin{figure}[t]
\centerline{
\includegraphics[trim=17mm 0mm 0mm 0mm,height=0.35\textheight,clip]{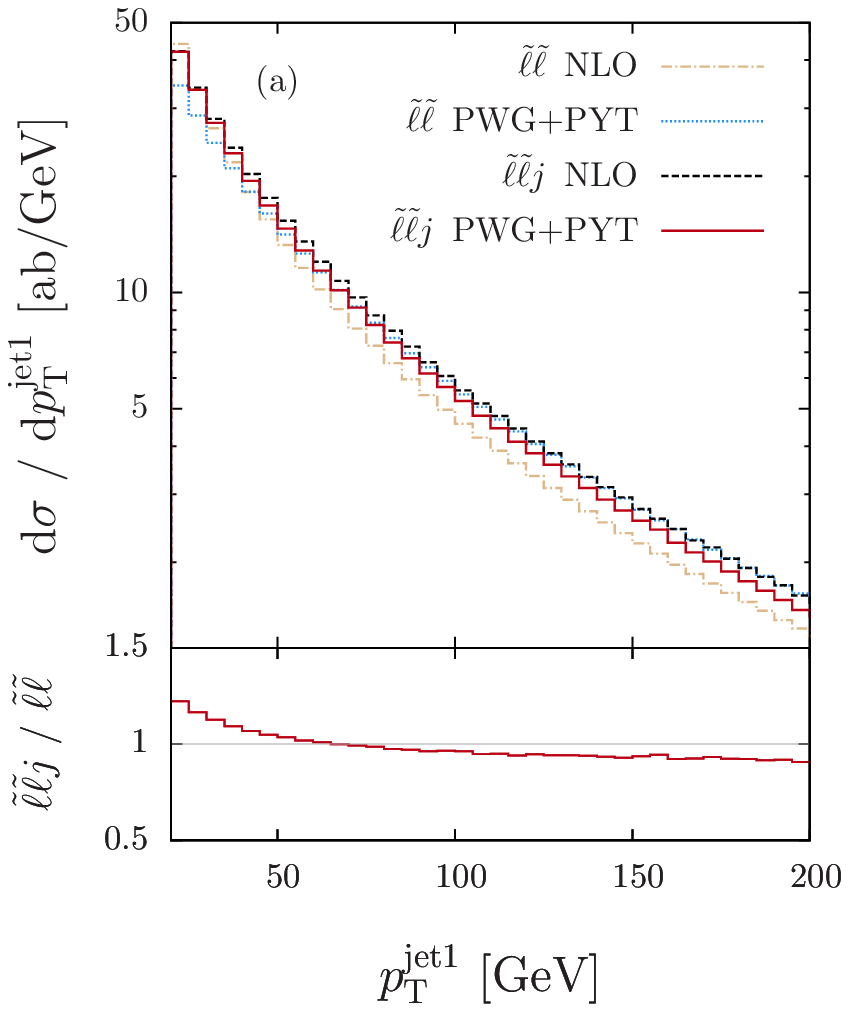}
\includegraphics[trim=12mm 0mm 0mm 0mm,height=0.35\textheight,clip]{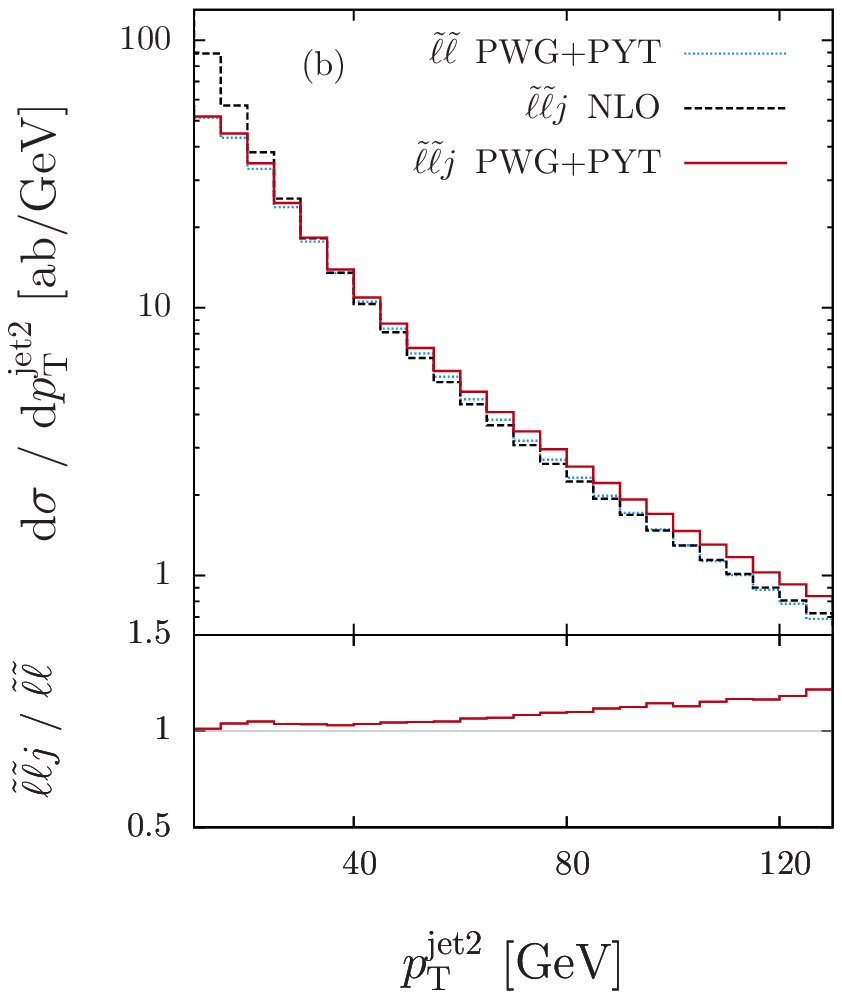}
}
\caption{\label{fig:ptjet}
Transverse momentum of the hardest jet (a) and the second-hardest jet (b) for our default setup in  $pp\to \slsl+X$ at NLO (beige dot-dashed), $pp\to \slsl+X$ with \POWHEGpPYTHIA{} (blue dotted), $pp\to \slslj+X$ at NLO (black dashed), and $pp\to \slslj+X$ with \POWHEGpPYTHIA{} (red solid). 
The respective ratios of the \POWHEGpPYTHIA{} results for 
$pp\to \slslj+X$ and $pp\to \slsl+X$ are shown in the lower panels.
}
\end{figure}

\begin{figure}[t]
\centerline{
\includegraphics[trim=17mm 0mm 0mm 0mm,height=0.35\textheight,clip]{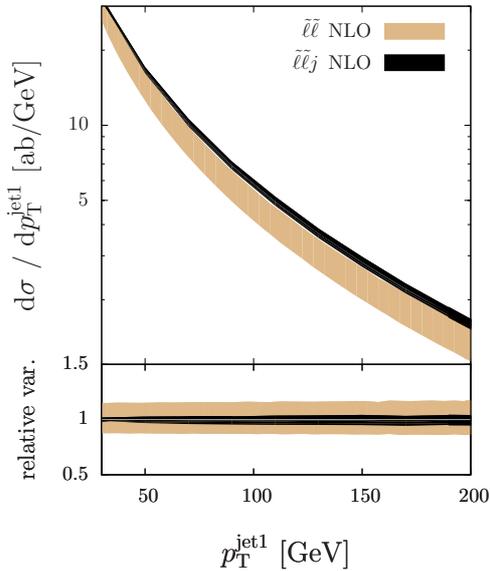}
}
\caption{\label{fig:ScaleVar}
Scale variation for the transverse momentum distribution  of the hardest jet at NLO  for $pp\to \slsl+X$ (beige) and $pp\to \slslj+X$ (black) with $0.5\mu_0\leq \mu_\mr{R},\mu_{F}\leq 2\mu_0$ (upper panel). The respective variations relative to the central scale choice $\mu_0$ are given in the lower panel. 
}
\end{figure}

In  Ref.~\cite{Jager:2012hd} we have investigated inclusive slepton pair production, $pp\to \slsl+X$, at NLO-QCD, matched with \PYTHIA{} via the \POWHEG{} approach. In that work, distributions related to the hardest jet could be accounted for only at leading-order, whereas our new implementation of $pp\to \slslj+X$ provides full NLO-QCD accuracy for this class of observables.  The transverse momentum distributions of the hardest jet in the two approaches are shown in Fig.~\ref{fig:ptjet} (a).
Scale uncertainties in this jet distribution, obtained by varying the factorization and renormalization scales independently in the range $0.5\mu_0\leq \mur,\muf\leq 2 \mu_0$, are provided in Fig.~\ref{fig:ScaleVar}.
The widths of the bands can be considered as rough indicators for the theoretical uncertainties of the respective predictions. Clearly, providing matrix elements at NLO-QCD accuracy for $\slslj$ final states much improves the uncertainty associated with jet distributions emphasizing the need for such a calculation, in particular for observables related to the jet. 

While in $pp\to \slsl+X$ any jet apart from the hardest can be simulated by the parton shower only, in $pp\to \slslj+X$ a second hard jet can be accounted for by the real-emission contributions of the hard matrix elements. Figure~\ref{fig:ptjet} (b) shows the transverse momentum distribution of the second-hardest jet as obtained with the respective programs for $pp\to \slsl+X$ and  $pp\to \slslj+X$. Clearly, the inclusion of contributions from the matrix element in $pp\to \slslj+X$ gives a better description of hard jet configurations, while the Sudakov factor provided by the \POWHEGpPYTHIA{} implementation provides the expected suppression of contributions with low transverse momenta.

\begin{figure}[t]
\centerline{
\includegraphics[trim=17mm 0mm 0mm 0mm,height=0.25\textheight,clip]{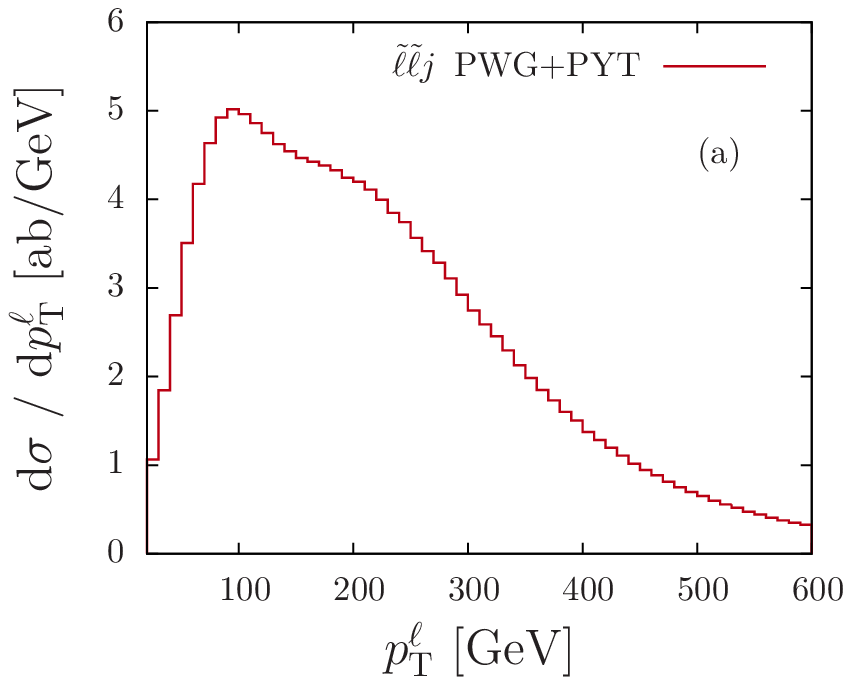}
\includegraphics[trim=8mm 0mm 0mm 0mm,height=0.25\textheight,clip]{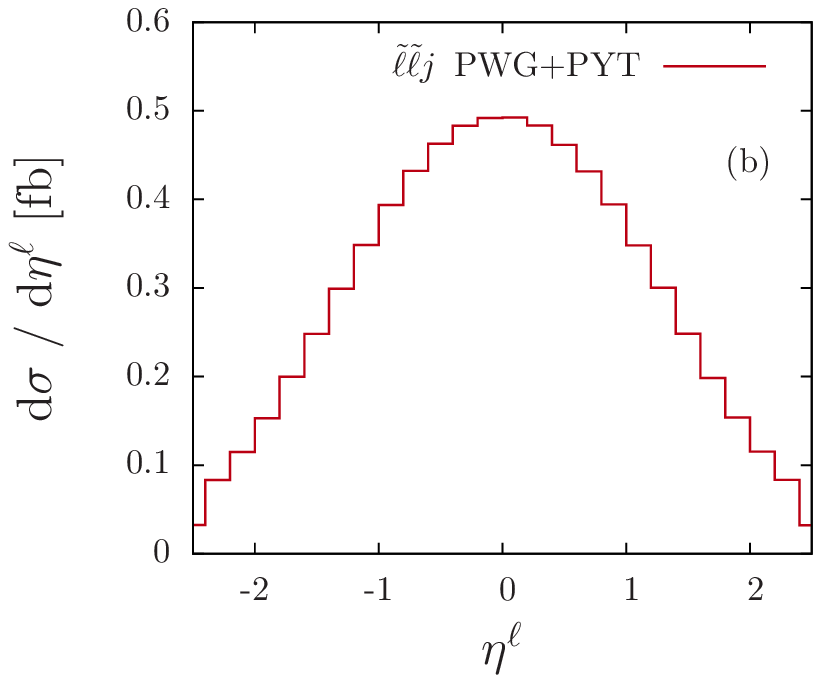}
}
\caption{\label{fig:lepton}
Transverse momentum (a) and rapidity distribution (b) of the hardest negatively charged lepton in $pp\to \slslj+X$ (red solid) with \POWHEGpPYTHIA{}. 
}
\end{figure}

A realistic analysis of slepton pair production processes requires access to the kinematic properties of the decay products of the heavy SUSY particles. Such decays can be conveniently simulated by {\PYTHIA}. 
To illustrate the capability of our code that handles decays of the sleptons via an interface to \PYTHIA{}, we consider a setup with two oppositely charged hard, central leptons,
\begin{equation}
\label{cuts:lepton}
p_\mr{T}^\ell > 20~\mr{GeV}\,,
\quad
|\eta^\ell|<2.5\,,
\end{equation}
in the presence of a hard jet, fulfilling Eq.~(\ref{cuts:basic}). The leptons are required to be well-separated from each other and from the jets, in the rapidity-azimuthal angle plane, 
\begin{equation}
\label{cuts:delta-r}
\Delta R_{\ell\ell} > 0.4\,,\quad
\Delta R_{\ell j} > 0.4\,.
\end{equation}
In Fig.~\ref{fig:lepton}
we show the transverse momentum and rapidity distributions of the hardest negatively charged lepton in $pp\to \slslj+X$ as obtained with \POWHEGpPYTHIA{}.

Since we are having full access to the kinematics of the decay leptons, we can also provide distributions for sophisticated observables that are typically used in SUSY analyses for an optimal signal selection in the presence of background processes with a priori large event rates. 
In Ref.~\cite{Buckley:2013kua}, {\em super-razor variables} have been introduced as a means to improve searches for weakly interacting new particles that are produced in pairs at the LHC, such as charginos and sleptons.  Super-razor variables are constructed by approximate boosts to the center-of-mass frame of the slepton-pair system, followed by boosts to the slepton decay frames, see Ref.~\cite{Buckley:2013kua} for details. In the presence of QCD radiation,
boosts against the jet directions are included. Due to this construction, extra jets in an event do not alter the shape of super-razor variables. The $\slsl+X$ and $\slslj+X$ processes thus exhibit super-razor variables of the same shape.

\begin{figure}[t]
\centerline{
\includegraphics[trim=17mm 0mm 0mm 0mm,height=0.25\textheight,clip]{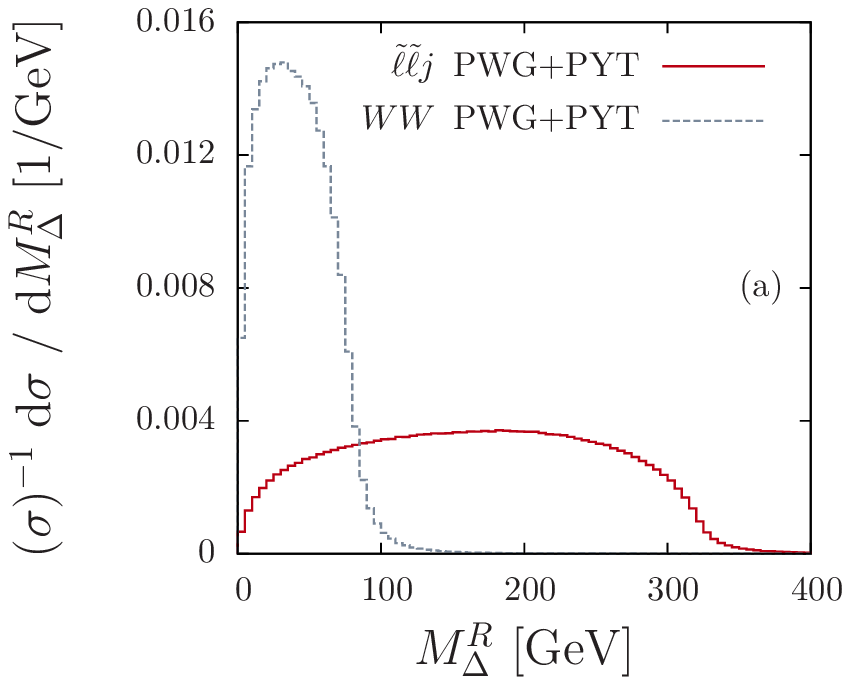}
\includegraphics[trim=12mm 0mm 0mm 0mm,height=0.25\textheight,clip]{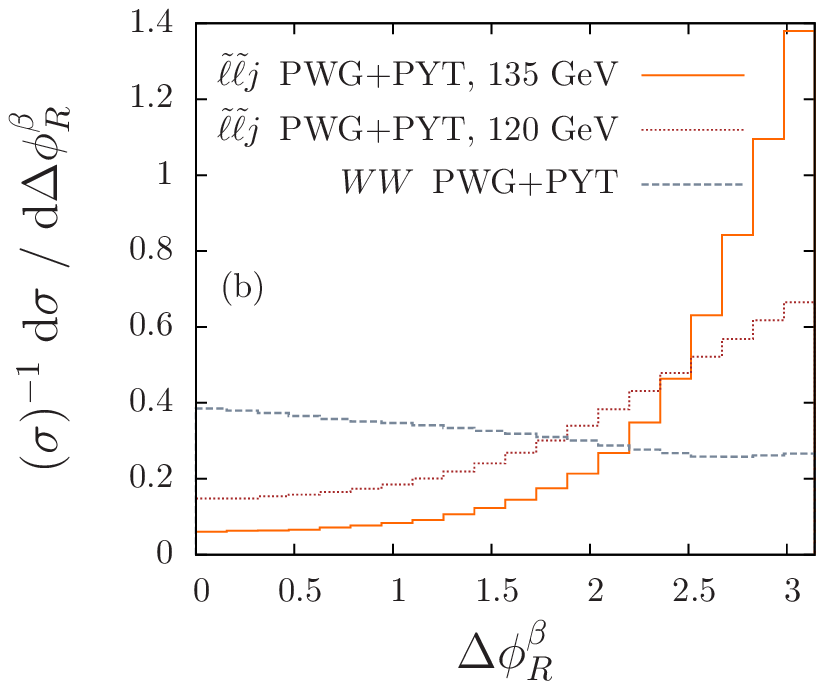}
}
\caption{\label{fig:SuperRazor}
Left panel: normalized distribution of the super-razor variable $M^R_{\Delta}$ in our default MSSM setup,
$m_{\tilde{\ell}} = 350\text{~GeV}$,
for $pp\to \slslj+X$ (red solid) and for $pp\to W^+ W^- +X$ (grey dashed). 
Right panel: normalized distribution of the super-razor variable $\Delta \phi^{\beta}_R$ for $pp\to \slslj+X$ in a light slepton setup,
$m_{\tilde{\ell}} =  150\text{~GeV}$,
with $m_{\tilde{\chi}_1^0}=135$~GeV (orange solid), $m_{\tilde{\chi}_1^0}=120$~GeV (brown dotted) and for $pp\to W^+ W^- +X$ (grey dashed). 
}
\end{figure}

The super-razor variable  $M_\Delta^R$ contains information about the mass differences involved in the pair production and subsequent decay. 
For slepton pair production processes, the $M^R_{\Delta}$ distribution drops rapidly at $M^R_{\Delta} = M_\Delta$, with
\begin{equation}
M_{\Delta} = \frac{m_{\tilde \ell}^2 - m_{\tilde \chi}^2}{m_{\tilde \ell}}.
\end{equation}
Background processes do not contain information about that mass scale, and thus do not exhibit a pronounced edge at the value of $M^R_{\Delta}$ characterized by the mass difference between slepton and neutralino. 
Applying the cuts of Eqs.~(\ref{cuts:basic}), (\ref{cuts:lepton}), (\ref{cuts:delta-r}), in Fig.~\ref{fig:SuperRazor} (a)
we show the $M^R_{\Delta}$ distribution of the $pp\to \slslj+X$ signal process in comparison to the 
$pp\to W^+ W^- +X$ background, restricting ourselves to decays of the $W$-bosons into first-generation lepton-neutrino pairs. The results for the background process we have obtained with the help of the corresponding \POWHEGBOX{} implementation~\cite{Melia:2011tj}.

Additional information on the kinematics of the reaction is provided by angular variables, such as $\Delta\phi_R^\beta$, that in the super-razor approach is constructed from the boost direction and the momenta of the visible decay particles.  Such angular variables are particularly powerful in scenarios where the mass difference of the sleptons and the neutralinos they decay into is small. For a mass difference of 50~GeV or less, sleptons as light as 100~GeV are not yet excluded. We study this case in a {\em light slepton setup} with a slepton mass of $m_{\tilde{\ell}} = 150$~GeV and different values of the lightest neutralino mass $m_{\tilde{\chi}_1^0}$ close to $m_{\tilde{\ell}}$ and, again, the cuts of  Eqs.~(\ref{cuts:basic}), (\ref{cuts:lepton}), (\ref{cuts:delta-r}). The discriminatory power of $\Delta\phi_R^\beta$ in the light slepton setup is illustrated in Fig.~\ref{fig:SuperRazor} (b) for neutralino masses of $m_{\tilde{\chi}_1^0} = 120$~GeV and $m_{\tilde{\chi}_1^0} = 135$~GeV, respectively. While for the signal 
process the leptons tend to be aligned with each other, opposite to the boost direction, resulting in a peak of the  $\Delta\phi_R^\beta$ distribution around $\pi$, especially for a small mass difference between slepton and neutralino, the $W^+W^-$ background does not exhibit such a correlation, but features a rather flat distribution.

\begin{figure}[t]
\centerline{
\includegraphics[trim=15mm 0mm 0mm 0mm,height=0.25\textheight,clip]{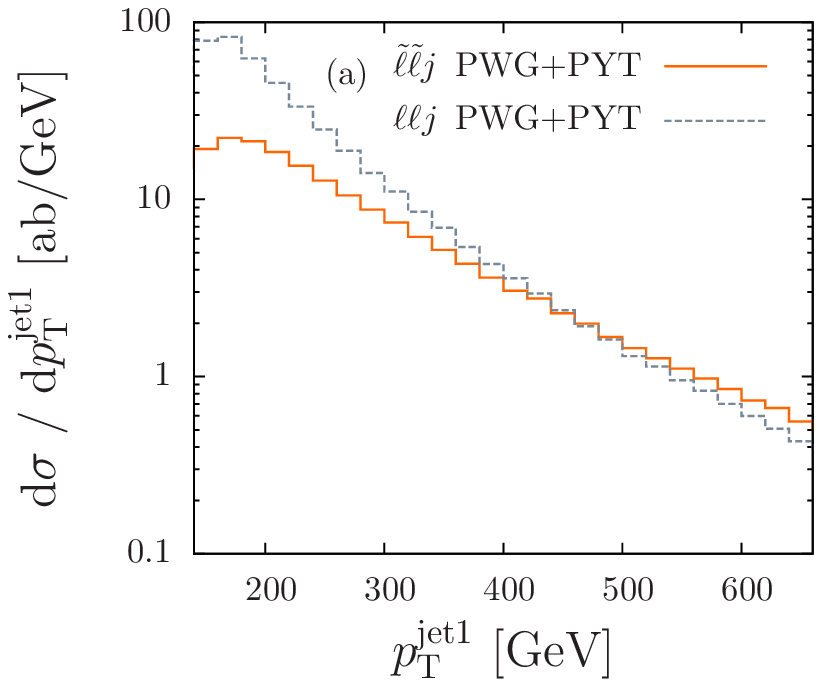}
\includegraphics[trim=15mm 0mm 0mm 0mm,height=0.25\textheight,clip]{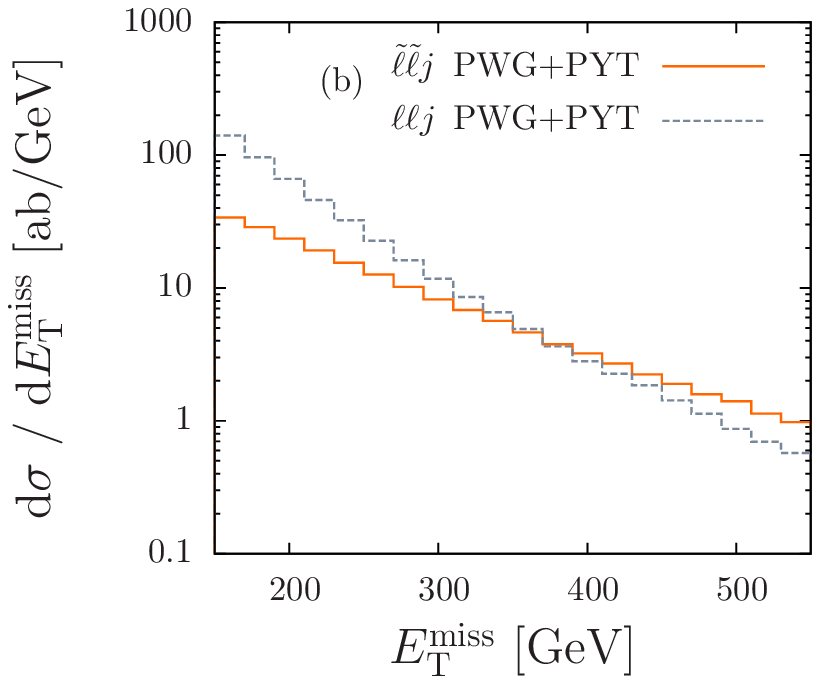}
}
\caption{\label{fig:monojet}
Transverse momentum of the hardest jet (a) and missing transverse energy (b) for $pp\to \slslj+X$ (orange solid) and $pp\to \ell^+\ell^-+\mr{jet}+X$ (grey dashed) in our monojet analysis setup.
}
\end{figure}
In SUSY scenarios where the masses of the sleptons and neutralinos are almost degenerate, slepton decays produce neutralinos and very soft leptons that easily escape detection. The tell-tale signature of such reactions is the missing transverse energy associated with the neutralinos accompanied by hard QCD radiation resulting in a monojet configuration. 
To make use of this signature, an accurate, matrix-element based description of the hard jet is mandatory. We illustrate the capability of our code for $pp\to\slslj+X$ to serve that purpose in the light slepton scenario for neutralinos with a mass of $m_{\tilde{\chi}_1^0}=135$~GeV, that are almost degenerate with the $m_{\tilde{\ell}} = 150$~GeV sleptons.

Following the strategies for monojet searches as a tool for the discovery of only weakly interacting SUSY particles at hadron colliders presented in \cite{Gunion:1999jr,Han:2013usa,Schwaller:2013baa,Baer:2014cua}, we require at least one hard jet,
\begin{equation}
\label{cuts:monojet-jet}
p_\mr{T}^\mr{jet1}>120~\mr{GeV}\,,\quad
|y^\mr{jet1}|<4.5\,,
\end{equation}
and large missing energy, computed from all observed tracks in an event, 
\begin{equation}
\label{cuts:monojet-miss}
E_T^\mr{miss}>150~\mr{GeV}\,. 
\end{equation}
The cuts of Eqs.~(\ref{cuts:monojet-jet})--(\ref{cuts:monojet-miss}) are constructed to account for efficiency requirements in the missing energy triggers of the LHC experiments and to suppress background contributions that are a priori dominant at low transverse momenta, such as weak boson production in association with a jet.
To sketch the general features of such background processes,  we use the \POWHEGBOX{} implementation~\cite{Alioli:2010qp} for $pp\to Z+\mr{jet}+X$ in the $Z\to e^+ e^-$ decay mode, assuming that the decay leptons escape detection and give thus rise to missing transverse energy. 
Figure~\ref{fig:monojet} shows the transverse momentum and the missing transverse energy distributions of the signal and background processes within the cuts of Eqs.~(\ref{cuts:monojet-jet})--(\ref{cuts:monojet-miss}).  In each case, 
the signal contribution takes over in the tail of the distribution, thus confirming the desired impact of hard transverse momentum cuts on the signal significance.
We note that a full signal-to-background analysis would require considering all possible decay modes of the $Z$ boson and, moreover, additional background processes such as $pp\to W+\mr{jet}+X$. Such a detailed analysis	is, however, far beyond the scope of this work,
where we intend to simply illustrate the general benefit of monojet analyses in the context of slepton pair production processes
and point out the usefulness of our code for such studies.


\section{Summary and conclusions}
\label{sec:conc}
In this work we have presented an NLO-QCD calculation for slepton pair production in association with a hard jet at the LHC, and its matching with parton-shower programs in the framework of the \POWHEGBOX{}. While the reader is free to download the publicly available computer package and use it for applications of his own, we have presented numerical results for selected  phenomenological applications to illustrate the capability of our code and demonstrate the impact of radiative corrections and parton-shower effects on realistic analyses. 

We found that the NLO-QCD corrections to $pp\to\slslj+X$ are sizable, thus crucially requiring hard matrix elements at order $\mathcal{O}(\alpha_s^2\alpha^2)$ in order to ensure sufficient accuracy for cross sections and distributions, in particular for observables that are sensitive to the emission of hard jets. 
The matching of the NLO calculation with \PYTHIA{} allows us not only to account for -- generally small -- parton-shower effects, but also provides us with a convenient tool for the simulation of slepton decays. This feature puts us into a position to compute observables constructed from the momenta of decay products, such as the so-called super-razor variables. 
Having full control on the hard jet in $pp\to\slslj+X$, we can also provide reliable predictions for monojet analyses which have been developed for the extraction of SUSY signatures that are difficult to detect by other means.

%
\acknowledgments{
We are grateful to Jos{\'e} Zurita for many valuable discussions.
We thank Stefan Berge and Stefan Weinzierl for clarifying discussions,
and Thomas Hahn and Emanuele Re for useful comments.
This work was supported in part by the Institutional Strategy of the University of T\"ubingen (DFG, ZUK~63), by the GRK {\em Symmetry Breaking} (DFG/GRK 1581) of the German Research Foundation (DFG), the DFG Grant JA~1954/1, and by the PRISMA Cluster of Excellence.
}


\end{document}